\documentclass[journal,10pt]{IEEEtran}
\usepackage{amsmath}
\usepackage{bm}
\usepackage{subfigure}
\usepackage{graphicx,epsfig,balance}
\graphicspath{{./fig/}}
\usepackage{cite}
\usepackage{framed}
\usepackage{enumitem}
\usepackage{url}
\usepackage{soul}
\usepackage{color, xcolor}
\usepackage{multirow}
\usepackage{amsfonts,amssymb}
\usepackage{booktabs}
\usepackage{threeparttable}
\usepackage{soul}
\usepackage{listings}
\lstnewenvironment{queryl}[1][]
 {\lstset{frame=shadowbox,basicstyle={\small\ttfamily},escapechar=`,breaklines=true, linewidth=8cm, #1}}
 {}

\definecolor{dkgreen}{rgb}{0,0.6,0}
\definecolor{gray}{rgb}{0.5,0.5,0.5}
\definecolor{mauve}{rgb}{0.58,0,0.82}
\usepackage[framemethod=TikZ]{mdframed}
\usepackage{lipsum} 

\newmdenv[
  linecolor=black,
  linewidth=1pt,
  roundcorner=5pt,
  backgroundcolor=black!5,
  innertopmargin=5pt,
  innerbottommargin=10pt,
  innerrightmargin=10pt,
  innerleftmargin=10pt,
  frametitlealignment=\centering,
]{examplebox}

\soulregister\cite7 
\soulregister\citep7 
\soulregister\citet7 
\soulregister\ref7 
\soulregister\pageref7 
\makeatletter
\newif\if@restonecol
\makeatother

\usepackage[linesnumbered,ruled,lined]{algorithm2e}
\usepackage{algpseudocode}
\usepackage{array}
\usepackage{makecell}

\usepackage{verbatim}
\usepackage{lettrine}

\newcommand{\qw}{{\bf w}}

\newcommand{\qh}{{\bf h}}
\newcommand{\qJ}{{\bf J}}
\newcommand{\qU}{{\bf U}}
\newcommand{\qa}{{\bf a}}

\newcommand{\qg}{\bm{g}}

\newcommand{\qzero}{\bm{0}}

\newcommand{\be}{\begin{equation}} \newcommand{\ee}{\end{equation}}
\newcommand{\bea}{\begin{eqnarray}} \newcommand{\eea}{\end{eqnarray}}

\begin{document}
	
\title{LLM-Enabled Automated Algorithm Design for Multiuser Fluid Antenna Communications}
	\author{
		 Gan Zheng,~\IEEEmembership{Fellow,~IEEE}, Fei Liu,  and Qingfu Zhang, ~\IEEEmembership{Fellow,~IEEE}
				\IEEEcompsocitemizethanks{
			\IEEEcompsocthanksitem G. Zheng is with the School of Engineering, University of Warwick, Coventry, CV4 7AL, UK (Email: gan.zheng@warwick.ac.uk).
\IEEEcompsocthanksitem F. Liu and Q. Zhang are with Department of Computer Science, City University of Hong Kong, Kowloon Tong, Hong Kong (Email: fliu36-c@my.cityu.edu.hk, qingfu.zhang@cityu.edu.hk).
	}}	
	
	\markboth{\textit{Accepted in IEEE Transactions on Wireless Communications}}
	{}
	\maketitle
	
	\begin{abstract}
Fluid antenna is a new reconfigurable antenna technology that can dynamically adjust the positions or ports of radiating elements and therefore provides a new degree of
freedom for wireless communications. However, the associated port selection is a challenging large-scale combinatorial optimization problem and difficult to solve.
Existing manually designed heuristic algorithms are not only labor-intensive, but cannot achieve satisfactory performance. In this paper, we propose a novel paradigm that
leverages large language models (LLMs)  for automated design of optimization algorithms for fluid antenna systems without manual hyperheuristic tuning.
Specifically, we study the problem of maximizing the minimum signal-to-interference-plus-noise ratio (SINR) in the downlink to ensure fairness among users by
optimizing port selection and beamforming. We investigate two LLM-enabled algorithm optimization strategies. The first is to optimize the crossover and mutation operations
to enhance the performance of the well-known genetic algorithm and the second is to design AutoPort, a new heuristic from scratch by LLM, to solve the optimization problem. Simulation results verify that the proposed method can achieve near-optimal performance and   significant improvement over the conventional genetic algorithm and the deep learning approach.
	\end{abstract}

	\begin{IEEEkeywords}
		Fluid antenna, port selection, large language models, evolution of heuristic, automated heuristic design
	\end{IEEEkeywords}
	
	\IEEEpeerreviewmaketitle
	
 \section{Introduction}
Multiple-input multiple-output (MIMO) technology has been a cornerstone for enhancing performance of wireless communications over several decades. However, the gains of MIMO come at a cost. Take massive MIMO in the fifth-generation (5G) mobile communications as an example.  The deployment of massive MIMO antenna arrays entails significant capital expenditures, including the cost of numerous antennas, space, radio frequency (RF) chains and energy consumptions.  It is therefore desirable to develop new cost-effective antenna solutions that can achieve performance comparable to massive MIMO, but without incurring its deployment and operational costs.

Recently, a novel reconfigurable antenna architecture known as the fluid antenna system (FAS) has   been proposed. FAS broadly encompasses any software-controlled fluidic, dielectric, or conductive structures, such as liquid-based antennas \cite{surfacewave}, pixel-based antennas \cite{pixel}, and metasurfaces \cite{meta} that are capable of dynamically reconfiguring their position, orientation, and other radiation characteristics. This reconfigurability enables the realization of spatial diversity within a physically constrained environment. The concept of FAS was first introduced to the wireless communications domain by Wong et al. in \cite{FAS-1}.  Since then, it has  attracted significant interests to develop efficient FAS for the sixth-generation (6G) mobile communications systems.
The fundamental concept of FAS involves the dense deployment of a large number of antenna ports within a limited physical space, while utilizing only a single RF chain \cite{FAS-2} and therefore substantially reduces the costs. The diversity analysis has revealed tremendous gain (more than 10x)  of FAS-enabled receivers  \cite{10130117} and  the diversity-multiplexing tradeoff has been
characterized from an information-theoretic perspective in \cite{10303274}.

FAS has been exploited together with the MIMO technology to boost the performance of both single-user and multiuser systems by leveraging conventional antenna
 arrays and flexible  port positions. However, MIMO-FAS also introduces new optimization challenges. The port selection is a high-dimensional
  combinatorial problem and is coupled with the traditional beamforming optimization, resulting in a difficult large-scale and highly non-convex problem.
  There have been various port selection methods for FAS \cite{port_2022,port_2024}. When both the MIMO transmitter and receiver have fluid antennas,
  quantum computation was studied in \cite{MIMO_port_CIM} and reduced exhaustive search and alternating optimization based on joint convex relaxation
  were proposed in \cite{MIMO_port_appro} to maximize the channel capacity. For the optimization of multiuser MIMO-FAS, alternating optimization techniques
   that iteratively optimizes port position and beamforming while keeping the other fixed,  is often employed \cite{fas_energy}\cite{fas_ISAC}. However,
   these methods not only incur  high complexity when solving subproblems at each iteration, but also cannot guarantee satisfactory performance.

Deep learning  has emerged as a new tool for optimization of MIMO-FAS.  Deep reinforcement learning has been applied to port selection and beamforming with promising performance gains \cite{DL1,DL2}. Graph neural networks (GNNs) have been studied which adopt a two-stage framework to solve the two subproblems of optimizing port position and beamforming separately \cite{DL5}, but this decoupling may ignore the relation between the two subproblems. More recently, large pre-trained large language models (LLMs) have been introduced to address the mobility issue in MIMO-FAS \cite{PS6}.
It utilizes LLMs as a pre-trained neural network together with low-rank adaptation  as a fine-tuning method to predict future channel state information (CSI) from historical CSI, and then select ports with higher accuracy in medium-to-high mobility scenarios. Intelligent FAS empowered by LLM has been proposed in \cite{LLM-Chao} where LLM has been applied for channel extrapolation, channel prediction, precoder design and autonomous FAS cooperation.
Nevertheless, deep learning based methods still face the challenges of large training data requirement, high training complexity, tedious parameter tuning, poor generalization and limited explainability for  MIMO-FAS.

Recently, there has been a new paradigm that leverages LLM to automatically design heuristic algorithms for various applications~\cite{llm_survey}. This approach utilizes an LLM's ability to comprehend algorithms and generate code to iteratively create heuristics, significantly reducing human design efforts. Unlike black-box deep-learning methods, the resulting heuristics are interpretable. For example, evolution of heuristics (EoH)~\cite{eoh,liu2023algorithm} uses an evolutionary framework to refine a population of heuristics with an LLM. Each heuristic is defined by its conceptual idea and code implementation. The LLM generates new or improved heuristics, which are then evaluated on target problem instances. The population is subsequently updated by retaining the best-performing heuristics. This cycle of generation, evaluation, and selection drives the automated discovery of effective heuristic algorithms. Similarly, Funsearch~\cite{romera2024mathematical} adopts LLMs in an evolutionary search framework for automated function search for mathematical discoveries. Instead of using a simple population management method, a multi-island approach is adopted to enhance diversity. This paradigm has been extended and applied to diverse optimization algorithm design tasks, including routing~\cite{eoh}, scheduling~\cite{li2025llm}, integer linear programming~\cite{ye2025large}, and Bayesian optimization~\cite{yao2024evolve}.

 This paper proposes an LLM-enabled framework for the optimization of multi-user FAS downlink multiple-input-single-output (MISO) communications systems, with the objective to maximize the minimum   signal-to-interference-plus-noise ratio (SINR)
 across users under the total transmit power constraint. The optimization problem entails port selection and the corresponding beamforming design. While given the port selection, an efficient algorithm to optimize the beamforming is available in the literature \cite{sinr_balancing}, the optimization of port selection
 is a challenging problem. Different from existing methods in the literature that manually design algorithms to optimize port selection and beamforming directly, our proposed method searches for
 heuristic algorithms automatically to solve the problem. In other words, we optimize algorithms instead of variables. Our key contributions  are summarized as follows:
	\begin{enumerate}
		\item This paper presents the first LLM-enabled algorithm design framework  for  SINR maximization in multi-user FAS downlink multiple-input-single-output (MISO) communications systems. This framework addresses limitations of existing methods, which automates the creation of customized and fully interpretable heuristic  algorithms without domain knowledge, pre-trained data or parameter tuning. It not only minimizes manual efforts, but also  leads to superior performance and potentially brings insights on new algorithms design.

 Note that our main purpose is to design customized heuristic algorithms well suitable for the SINR maximization problem which cannot be easily achieved by manual design and hyper-heuristic search, but the proposed framework does not always necessitate a fundamentally new algorithm.
		\item We propose two design strategies to solve the port selection problem. The first is based on the genetic algorithm (GA), and we utilize LLM to optimize the crossover and mutation functions. The second is AutoPort, completely designed by prompting LLMs, which adopts an evolutionary search framework, iteratively searching for algorithms that lead to higher SINR.
		\item Extensive simulations demonstrate the effectiveness of our proposed method. Results show that the first strategy can significantly improve the performance over the baseline GA while the second strategy can achieve even better and near-optimal user performance  at the cost of more running time.
	\end{enumerate}

 Our contributions go beyond a straightforward application of existing LLM frameworks because we reformulate and simplify the original sophisticated problem
to be suitable for the automated algorithm search, propose both incremental and discovery design paths, and   then customize the tasks and design task-specific prompts to be  tightly integrated into the heuristic search process by LLMs.

The remainder of this paper is organized as follows. Section II introduces the multiuser FAS downlink system model  and  details of the optimization problem. Section III presents the preliminaries of  LLM-enabled automated algorithm design. Section IV introduces details of the two design strategies to enhance the GA and design a new heuristic algorithm.
Section V provides numerical evaluations and performance comparisons. Finally, concluding remarks and discussions on future directions are given in Section VI. 	
	
\textit{Notation:} Bold lowercase and uppercase letters denote vectors and matrices, respectively. Superscripts $(\cdot)^T$ and $(\cdot)^\dag$ represent the transpose and Hermitian transpose. $\mathbb{C}^{m \times n}$ and $\mathbb{R}^{m \times n}$ denote the sets of $m \times n$ complex- and real-valued matrices, respectively.  The notation $\mathcal{CN}(\boldsymbol{\mu}, \boldsymbol{\Gamma})$ denotes a circularly symmetric complex Gaussian distribution with the mean vector $\boldsymbol{\mu}$ and the covariance matrix $\boldsymbol{\Gamma}$.  $\|\qa\|$ denotes the Frobenius  norm of a vector $\qa$. $|\cdot|$ denotes the absolute value of a scalar or the cardinality of a set.

 \section{System Model and Problem Formulation}
  \begin{figure}[h]
    \centering
    \includegraphics[width=0.8\linewidth]{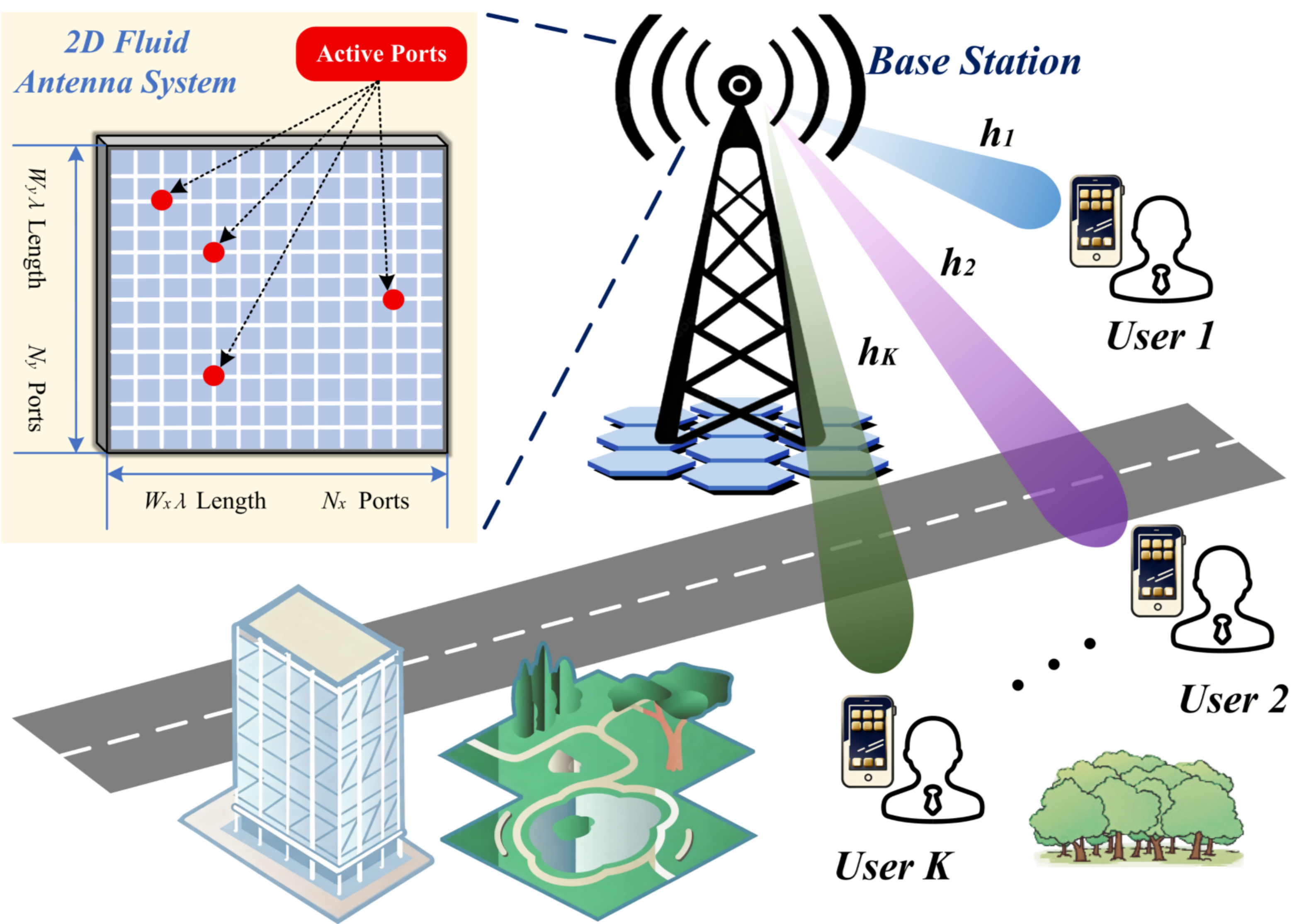}
    \caption{System model of a MISO downlink communications system using 2D-FAS at the BS.}
    \label{fig:sys}
 \end{figure}
We consider a multiuser MISO-FAS downlink system as illustrated in Fig. \ref{fig:sys}, in which the base station (BS) equipped with a two-dimensional (2D)  FAS array serves $K$ users each with a single fixed antenna.
The 2D FAS array has a dimension of $W = \lambda W_x \times \lambda W_y$, where $\lambda$ denotes the carrier wavelength.
$\lambda W_x$ and $\lambda W_y$ are the array lengths in the $x$ and $y$ directions.  The array consists of $N  = N_x\times N_y$ ports uniformly distributed across the aperture, with $N_i$ ports aligned along each dimension $\lambda W_i$ for $i \in \{x, y\}$.  We assume the BS has $n$ RF chains and therefore only select or activate $n$ out of $N$ ports to serve $K$ users and it is required that $n\ge K$.

To characterize the correlation between any two ports in the 2D array, we assume a rich scattering environment and   employ
the classical Jake’s model for the correlation  \cite{10303274}\cite{Jakes} \footnote{{{Note that other channel models may  apply such as the field-response
model    using the amplitude, phase, and angle of arrival/angle of departure information on
each  channel path under the far-field condition \cite{field_model}.}}}. Specifically, the spatial correlation between the $(n_x,n_y)$-th and the $(n_i,n_j)$-th ports is expressed as
\bea\label{eqn:correlation}
    &&J_{(n_x,n_y), (n_i,n_j)} = J_0\left(\frac{2\pi d}{\lambda}\right),\notag \\
     &=& J_0\left( 2\pi \left(\frac{|n_x-n_i|}{N_x-1}   W_x \right)^2 + \left(\frac{|n_y-n_j|}{N_y-1}    W_y\right)^2 \right),
\eea
where $d = \sqrt{ \left(\frac{|n_x-n_i|}{N_x-1} \lambda W_x \right)^2 + \left(\frac{|n_y-n_j|}{N_y-1} \lambda  W_y\right)^2 }$ denotes the distance between these two ports,
and $J_0$ is the zero-order spherical Bessel function.

For notation convenience, we map the 2D port coordinate $({n_x,n_y})$  to the 1D port index as follows:
 \be\label{eqn:mapping}
    n_{n_x,n_y} = (n_y-1) N_x + n_x \in \{1,\cdots, N\}.
\ee
We can therefore construct the spatial correlation matrix $\qJ \in \mathbb{R}^{N\times N}$ of the 1D channel between any two ports using \eqref{eqn:correlation} and \eqref{eqn:mapping}.

Apparently the matrix $\qJ$ is  symmetric, and its eigenvalue decomposition is
\be
    \qJ = \qU \Lambda \qU^\dag,
\ee
where $\qU$ contains the eigenvectors and $\Lambda$ is a diagonal matrix and its diagonal elements are eigenvalues.

Then the channel from all antenna ports at the BS to   user $k$ can be expressed as
\be
     \qh_k = \sqrt{\rho_k}\qU\sqrt{\Lambda} \qg_k,
\ee
where ${\rho_k}$ denotes the large-scale path loss,  $\qg_k\in \mathbb{C}^{N\times 1}$ denotes the small-scale fading channel vector that follows the complex Gaussian distribution, i.e., $\mathcal{CN}(\qzero, \boldsymbol{I})$.

 We assume full CSI of all  ports is available which is necessary to determine the optimal configurations.
Full CSI can be acquired by CSI extrapolation techniques that exploit correlations and predict full CSI from partial observations \cite{csi}.

Suppose  the set of chosen ports is denoted as $\phi$ which contains $n$ non-repeated ports, and the resulting channel of user $k$ is $\qh_k(\phi)\in \mathbb{C}^{n \times 1}$. Denote the transmit signal and the beamforming vector for user $k$ as $s_k$ with unit power and $\qw_k$, respectively. The received signal at user $k$ is
\be
    z_k = \qh_k^T(\phi)\qw_k s_k + \sum_{j=1, j\ne k}^K \qh_k^T (\phi)\qw_j s_j + n_k,
\ee
where $n_k$ is the additive white Gaussian noise with zero mean and variance $\sigma^2$.
The  received SINR of user $k$ is
\be
    \gamma_k = \frac{|\qh_k^T (\phi)\qw_k|^2}{ \sum_{j=1, j\ne k}^K |\qh_k^T (\phi)\qw_j|^2 + \sigma^2}.
\ee
The design objective is to maximize the minimum SINR of all users subject to the total power constraint $P$ by optimizing the port selection and the beamforming vectors. Mathematically, the problem is formulated as
\bea\label{eqn:P0}
    \max_{\{\qw_k\},\phi} && \gamma  \\
     \mbox{s.t.} && \gamma_k\ge \gamma, \forall k, \\
      && \phi  \subset \{1, \cdots, N\}, |\phi|=n, \\
     &&\sum_{k=1}^K \|\qw_k\|^2\le P.
\eea

Given $\phi$ or equivalent channels $\{\qh_k(\phi)\}$, the problem \eqref{eqn:P0} has been well studied and the solution to find the optimal $\{\qw_k\}$ is available
based on the uplink-downlink duality \cite{sinr_balancing}. Therefore in this paper we focus on the optimization of port selection $\phi$.

Assume given $\phi$ and the corresponding beamforming vectors, the optimal SINR is $\gamma(\phi)$ by using the algorithm in \cite{sinr_balancing}. Then the problem \eqref{eqn:P0}  can be simplified and rewritten as
\bea\label{eqn:P1}
      \max_{\phi} && \gamma(\phi)\\
      \mbox{s.t.} && \phi \subset \{1, \cdots, N\}, |\phi|=n.\notag
\eea
The above  selection problem is to choose $n$ out of $N$ ports to maximize the SINR.  If we use the exhaustive search to solve it, the number of all possible combinations is ${N \choose n}$, and
its complexity grows even faster than the exponential increase.
When $n$ or $N$ is large, the number of possible solutions is huge and cannot be enumerated. For instance, when $N=64$, the numbers of all possibilities for $n=4$ and $n=8$ are 635,376 and $4.4262\times10^{9}$, respectively. In this paper, we will leverage LLMs to design efficient heuristic algorithms to solve the problem \eqref{eqn:P1}.

 \section{Preliminaries: LLM-enabled Automated Algorithm Design}

 Recently, a new paradigm, often referred to as LLM-driven automated heuristic design (AHD), has emerged that leverages LLMs to automate the
 design of interpretable heuristic algorithms, significantly reducing the reliance on tedious hand-crafting and domain model training. At its core,
  this paradigm treats the heuristic itself as the target of an optimization search. Instead of searching for a solution to a problem instance,
   it searches the space of possible heuristics to find one algorithm that performs well across many problem instances. To formalize this, we can define the AHD process as follows.

 \begin{figure}[h]
    \centering
    \includegraphics[width=1\linewidth]{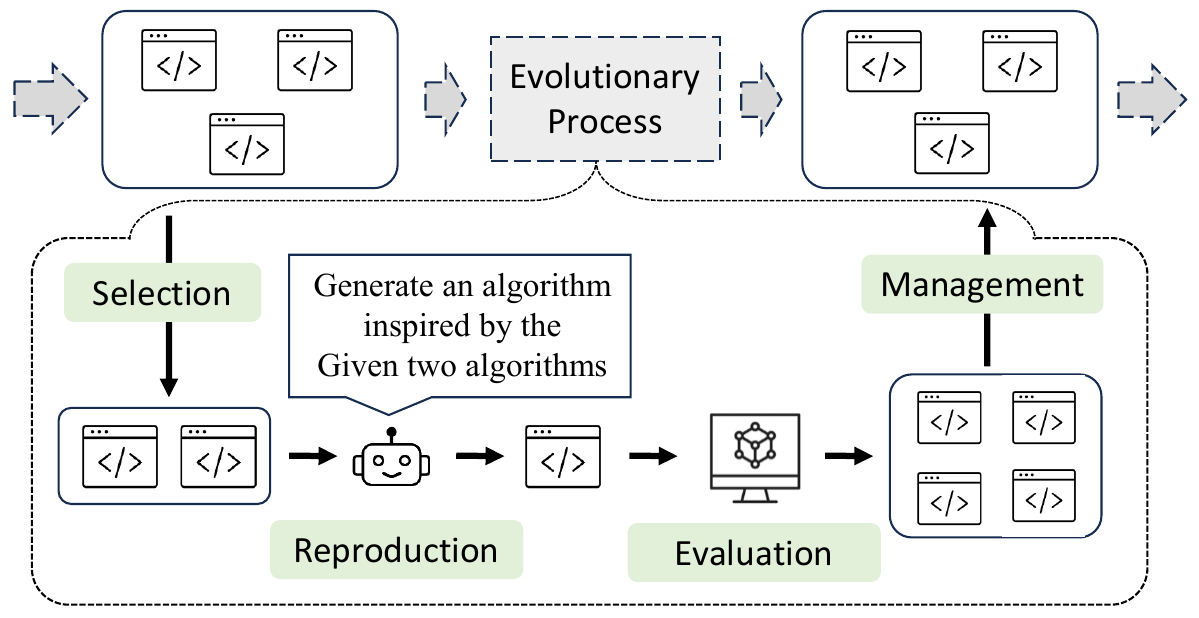}
    \caption{The pipeline of LLM-enhanced automated heuristic design. }
    \label{fig:pipeline}
 \end{figure}

Let the heuristic design task be denoted by $T$. For our paper, this task is to design a heuristic for the port selection problem defined in \eqref{eqn:P1}. We evaluate any candidate heuristic on a set of given problem instances, $I$, where each instance $i \in I$ corresponds to a specific channel setting. The goal is to find an optimal heuristic algorithm $a^*$ from a vast space of possible heuristics, $A^S$, that maximizes a performance metric, $F(a, I)$. In our context, this metric is the average minimum SINR achieved by a heuristic algorithm $a$ across all channel instances in $I$. This can be expressed as:
\begin{equation}
    F(a, I) = \frac{1}{|I|}\sum_{i \in I} f(a, i), \quad a \in A^{S},
    \label{eq:perf_metric}
\end{equation}
where $f(a, i)$ is the performance score (i.e., the resulting minimum SINR $\gamma(\phi)$) obtained by applying heuristic $a$ to instance $i$. The optimal heuristic $a^*$ is the one that solves the following optimization problem:
\begin{equation}
    a^* = \underset{a \in A^{S}}{\arg\max} \, F(a, I).
    \label{eq:opt_problem}
\end{equation}

Since the algorithm space $A^S$ is immense and unstructured, solving \eqref{eq:opt_problem} directly is intractable. AHD addresses this challenge by employing an evolutionary search framework powered by  LLM \cite{eoh}. Fig.~\ref{fig:pipeline} illustrates the general pipeline used by LLM-enabled AHD. In this framework, each heuristic $a$ is directly represented as a code implementation (e.g., Python implementation) and/or its natural language description~\cite{eoh}. The search begins with an initial ``population'' of algorithms and iteratively improves them over generations.
The process is summarized as follows:

\textbf{Step 1: Initialization}
An initial population of $m$ candidate algorithms, $P_0 = \{a_1, a_2, \dots, a_m\}$, is created. This is done by providing the LLM with a description of the optimization task $T$ and asking it to generate a diverse set of initial strategies.

\textbf{Step 2: Iterative Process}
To evolve the population from the $t$-th generation $P_t$ to the $(t+1)$-th population $P_{t+1}$, the following steps are repeated: 
\begin{itemize}
    \item \textbf{Step 2.1: Selection}
    ``Parent'' algorithms $a_p$ are selected from the current population $P_t$ according to certain criteria. While any selection method can be used in EoH, this paper employs a rank-based probabilistic selection. All algorithms in the current population are ranked according to their fitness. Algorithm $a_i$ is then randomly selected with probability $p_i  \propto 1/(r_i+m)$, where $r_i$ is its rank and $m$ is the population size. This approach gives higher-ranked (fitter) heuristics a greater chance of being selected as parents while still maintaining diversity by allowing lower-ranked heuristics to be chosen occasionally.

    \item \textbf{Step 2.2: Reproduction}
    New ``offspring'' algorithms $a_o$ are created from ``parent'' algorithms $a_p$. It is guided by specific instructions (prompts) given to the LLM. For example, the LLM might be prompted to ``combine the strategies of these two parent algorithms'' or ``suggest a modification to this algorithm to improve its performance''. This can be abstractly represented as:
    $$a_o = \text{LLM}(\text{Prompt}(a_p)).$$
    Here, \texttt{Prompt()} constructs the instructions and provides the parent algorithm(s) as context, and \texttt{LLM()} represents the LLM's inference process that generates the new offspring algorithm.

    An example prompt used for reproduction in the bin packing problem \cite{binpacking} is illustrated in the following box:

    \begin{examplebox}[frametitle={Example Prompt for Reproduction}]
    I'm designing a heuristic for the bin packing problem. The goal is to minimize the number of used bins. At each step, an item is placed into the bin with the maximum score.

    Here are two existing heuristics:
    \begin{itemize}[leftmargin=3em, itemsep=2pt]
        \item[No.1] Heuristic description: \\
        \texttt{\hspace*{1em} <Code for heuristic 1>}
        \item[No.2] Heuristic description: \\
        \texttt{\hspace*{1em} <Code for heuristic 2>}
    \end{itemize}
    Design a new heuristic inspired by, but different from, the ones provided.
    \begin{enumerate}[label=\arabic*., wide, labelindent=0pt, topsep=3pt]
        \item Identify the common idea behind the provided heuristics.
        \item Describe your new heuristic in one sentence.
        \item Implement it in a Python function \texttt{score(item, bins)} that returns the \texttt{scores}.
    \end{enumerate}
    \end{examplebox}

    \item \textbf{Step 2.3: Evaluation}
    The fitness $F(a_o, I)$ of each new algorithm $a_o$ is measured. This is a practical step where the algorithm's code is executed on the set of channel instances $I$. The average performance score, as defined in \eqref{eq:perf_metric}, determines the algorithm's fitness.

    \item \textbf{Step 2.4: Management}
    After generating $m$ new offspring algorithms,
    a new population $P_{t+1}$ is formed by selecting the fittest $n$ algorithms from the combined pool of the current population $P_t$ and the $m$ new offspring algorithms.  This survival-of-the-fittest mechanism ensures that effective algorithmic ideas are carried forward and refined in subsequent generations.
\end{itemize}

\textbf{Step 3: Termination}
The evolutionary search loop continues until a stopping criterion is met, such as a maximum number of generations or a predefined budget for LLM queries. The final output is the best-performing algorithm $\hat{a}^*$ discovered during the search.

The LLM-driven AHD paradigm offers the following distinct advantages over existing methods:
\begin{itemize}
    \item Unlike manual design \cite{maunal}, it automates the discovery process, reducing human effort and potentially uncovering novel, non-intuitive strategies.
    \item  Compared to classical hyper-heuristic design \cite{hyper}, it substantially reduces the number of required evaluations of candidate heuristics across many problem instances, and the corresponding computational resources.
    \item  Compared to deep learning methods, which produce ``black-box'' neural networks, the final output of AHD is a fully interpretable, human-readable algorithm (e.g., a Python function).
    This transparency allows for verification and analysis, enabling researchers to gain new insights into the problem structure.
    Furthermore, AHD does not require the extensive data collection and lengthy training time  associated with deep learning models, and generalizes well to unseen scenarios.
\end{itemize}

 A detailed comparison of evaluation budget,  time to solution, and reduced dependence on human priors is provided in Table \ref{table:compare}.
\begin{table}
  \centering
 \begin{tabular}{|p{0.2\linewidth}|p{0.2\linewidth}|p{0.2\linewidth}|p{0.2\linewidth}|}\hline
 Comparison Aspects & Classical Hyper-heuristic & Manual Tuning & LLM-Enabled (This Paper) \\ \hline
 Candidate heuristics evaluated & 1,000$\sim$10,000+ & 10$\sim$50 (human-guided) & $\sim$100 (from EoH framework) \\ \hline
 Human effort & Medium (design of representation, operators)& High (expert intuition, experimentation) & Low (prompt engineering, result interpretation) \\ \hline
 Time to solution & Days$\sim$weeks (compute-bound) & Weeks$\sim$months (human-bound) & Hours$\sim$days (LLM generation + evaluation)\\ \hline
\end{tabular}
  \caption{Comparison with hyper-heuristics and manual meta-heuristic tuning methods}\label{table:compare}
\end{table}

In this paper, our research utilizes the LLM4AD platform \cite{LLM4AD}, an open-source Python library we have recently developed  to advance research in LLM-driven algorithm design.
LLM4AD features a modular architecture that decouples high-level search strategies from the underlying language models and task-specific evaluations.
This platform includes modules for search methods, a unified LLM programme template, and a sandboxed environment for securely evaluating candidate algorithms.
LLM4AD is easy to use, and we only need to specify the latter two modules for a specific problem, i.e., the programme template and the evaluation function. It supports a wide range of applications with an extensive library of around 100 tasks across optimization, machine learning, and scientific discovery.


 \section{Proposed Optimization Strategies}
 In this section, we present two strategies to use  LLM for designing efficient algorithms to solve the problem \eqref{eqn:P1}. The first one optimizes parts of operations of a basic GA and thus enhance its performance, while the second one prompts the LLM to design a complete algorithm by itself.
 For implementation, we employ EoH available in the LLM4AD platform.

 \subsection{Optimization of GA}
 We first describe the principle of a basic GA, and then introduce LLM to optimize two components of its operations: crossover and mutation.

 \subsubsection{Basic GA}
 The GA is a heuristic algorithm widely used to solve constrained and unconstrained optimization problem that is inspired by the natural selection.

It iteratively refines a population of individual solutions. Briefly speaking, at each iteration, the GA selects individuals from the current population to be parents and uses them to produce the children or offsprings via operations including crossover and mutation for the next iteration.  This process is repeated until convergence, e.g., for a fixed number of iterations or until no further improvement is made, and hopefully the population will evolve to an optimal solution.

In this paper, the specific basic GA employed  to solve \eqref{eqn:P1} is summarized as follows.
\begin{itemize}
  \item Initialization. The algorithm begins with random initial population with $M$ (an even number) individual port selection solutions. Each solution contains $n$ non-repeated ports.
  \item Create new populations of solutions by repeating the following steps at each iteration until convergence.
    \begin{itemize}
      \item Population Selection.   It evaluates the $M$ individual port selection solutions by using the objective function in \eqref{eqn:P1}, and then select the best half $\frac{M}{2}$ individuals as parents. It also reserves a small portion, e.g., $Mp, (0<p<1)$ individuals of the best solutions, called elite individuals, that will be passed to the next iteration directly without any further processing.

  \item Crossover.   It generates $M- Mp$ children solutions from $\frac{M}{2}$   parents solutions. The basic crossover function is to split ports in  the parents solutions in half (assuming $n$ is an even number), randomly choose $M- Mp$ solutions from each half solutions (which could be repeated), and then combine them together to form  $M- Mp$ new solutions.

  \item Mutation.   It introduces random changes in the population after crossover. This helps maintain diversity in the population and explores new solution space.  The basic mutation function used is to uniformly sample each port selection in the population and then randomly change it to a different port.

  \item Validation.
  After the above operations, the obtained individual port selection solutions may contain repeated ports and are therefore invalid. This step will identify these solutions and replace repeated ports with randomly chosen non-repeated ones.

  \item Population merging.   This step will merge the above obtained population with the elite individuals to form a new population of size $M$ to be used for the next iteration.

  \item Evaluation. In this step, the performance of the population will be evaluated using the objective function in \eqref{eqn:P1} and the best will be stored for the current iteration.
  \end{itemize}
\item The best solution  across iterations will be retained as the solution to the problem \eqref{eqn:P1}.
\end{itemize}

  \subsubsection{Optimizing Crossover}
  It can be seen from the above description that in the basic GA, although the general purposes of crossover and mutation are clear,   their implementations can vary according to the problem definitions and there is no optimal
  implementation in general. This gives rise to the optimization of the specific implementation for crossover and mutation functions. In this subsection, we first keep the basic mutation function and all other
   parts of the basic GA fixed, and only optimize the crossover function using EoH provided in the LLM4AD platform.

  Specifically, for EoH to automate the process of creating optimized crossover function, we provide EoH the  programme template  in Python as shown in Listing \ref{lst:cross:prompt}.
  In the template, we need to include   details of the input and output parameters together with  a clear and concise task description.
  Then EoH will instruct the LLM to evolve the heuristic design using  the template, together with the task evaluate function based on the objective function of
  \eqref{eqn:P1} and channel data.

 \begin{lstlisting}[caption={The programme template for the crossover function.},captionpos=b,label={lst:cross:prompt}]
template_program = '''
def crossover(parents,  M):
    """
    Design a crossover function to be used in the genetic algorithm.
    Args:
     parents: 2D integer Numpy array of shape (n_parents, n).
     M: Number of output populations.
    Returns:
    offspring: 2D integer Numpy array of shape (M, n).
    """
'''
task_description = "Design a crossover function for a genetic algorithm. The function performs a genetic crossover function on parents to generate multiple offspring."
\end{lstlisting}

EoH will then iteratively refine  the population of the crossover function by using the LLM until the termination condition is met. When evaluating a candidate crossover function, the LLM4AD will run the basic GA algorithm introduced in the previous subsection with the candidate crossover function.

By checking the log of the algorithm evolution, we have found some typical sample crossover algorithms designed by EoH as shown in Table \ref{table:cross}. We can see that the evolution has explored a broad range of possible crossover algorithms.
 \begin{table}[]
\centering
\caption{Typical crossover algorithms attempted in the evolution.}
\label{table:cross}
\begin{tabular}{ |l|p{6cm}| }  \hline
  No. & Description \\  \hline
  1 &   The algorithm performs a uniform crossover with probability-based gene selection, where each gene in the offspring is independently chosen from either parent with equal probability, creating highly diverse combinations while maintaining genetic material from both parents.\\ \hline
  2 &   The algorithm performs a dynamic segment swap crossover by randomly varying segment lengths and using a probabilistic swap threshold to create offspring, enhancing exploration while maintaining genetic diversity. \\ \hline
  3 &   The algorithm performs a circular shift crossover by rotating parent arrays at random pivot points and combining segments from the rotated parents to create offspring, introducing rotational diversity in recombination.  \\ \hline
 Chosen & The algorithm performs a gene frequency-based crossover with probabilistic inheritance, where each gene in the offspring is selected from either parent based on the relative frequency of that gene's value across the entire parent population.  \\ \hline
\end{tabular}
\end{table}
The finally chosen crossover function that gives the best performance for the training data is also shown in Table \ref{table:cross} and the Python code   is given in  Listing \ref{lst:crossover} of Appendix A.

 We can see that the  optimized crossover function adopts a frequency-based probabilistic inheritance strategy, which is an advanced combination technique.
Unlike uniform crossover, these methods leverage statistical data to bias inheritance toward beneficial traits. To be specific, whether a port is chosen or not in the offspring  is not determined randomly but by the relative frequency of that port's value across the entire parent population. The advantages of this crossover algorithm include
maintaining diversity while avoiding local optima and accelerating the search for the optimal solution.

  \subsubsection{Optimizing Mutation}
 To further enhance the performance of GA, next we optimize the mutation function to explore the solution space while using the above optimized crossover function.
 The programme template   in Python for LLM to automate the evolution of the mutation function is provided in Listing \ref{lst:mutate:prompt} below. Similar to the crossover function, the template describes the input and output parameters, as well as the task.

  \begin{lstlisting}[caption={The programme template for the mutation function.},captionpos=b,label={lst:mutate:prompt}]
template_program = '''
def mutation(population, N):
    """
    Design a mutation function to be used in the genetic algorithm.
    Args:
        population: Numpy array of shape (p, n).
        N: The largest integer for the population is N-1.
    Returns:
        mutated_population: Numpy array of shape (p, n).
    """
'''
task_description = "Design a mutation function for a genetic algorithm.  The function modifies a given 2D population array parents to ensure exploration of the genetic algorithm."
\end{lstlisting}

By checking the log of the algorithm evolution, we have found some typical sample mutation algorithms designed by EoH as shown in Table \ref{table:mutate}. Again we can see that the evolution has explored a broad range of possible mutation algorithms.
 \begin{table}[]
\centering
\caption{Typical mutation algorithms attempted in the evolution.}
\label{table:mutate}
\begin{tabular}{ |l|p{6cm}| }  \hline
  No. & Description \\  \hline
  1 &  The mutation function randomly selects a subset of genes in each individual and replaces them with new random values within the allowed range to ensure exploration.

  \textit{\textbf{Comment}}: This is the start of the evolution and is very much like the basic algorithm.   \\  \hline
  2 &  The mutation function applies a random cyclic shift to each row of the population array, preserving all elements but rearranging their positions. \\  \hline
  3 &  The new algorithm introduces diversity by flipping a variable number of bits in each individual, where the number of bits flipped is randomly chosen between 1 and 20\% of the individual's length, and the mutation values are drawn from a Gaussian distribution centered around the original values with a standard deviation of $\frac{N}{4}$.
\\ \hline
Chosen & The new algorithm introduces diversity by swapping pairs of elements in 50\% of individuals with 70\% probability, then applying uniform random noise to 15\% of elements in each individual, scaled by $\frac{N}{5}$ and rounded to integers. \\ \hline
\end{tabular}
\end{table}

%
%
The finally chosen mutation algorithm that gives the best performance for the training data is also shown in Table \ref{table:mutate} and the Python code   is  given
in Listing \ref{lst:mutate} of Appendix B.

 As can be seen, the key features of the chosen mutation algorithm for port selection include diversity and exploration. Rather than simply modifying the port selection in a uniform way in the traditional mutation operation,  the optimized mutation function  first introduces diversity by swapping pairs of elements in 50\% of individual solutions with 70\% probability. To encourage exploration, it then adds random noise that follows the uniform distribution $[-\frac{N}{5},\frac{N}{5}]$   to 15\% of port selections in each individual solution, and finally rounds them to integers. We have observed that the key parameters in this strategy, including the number of mutated solutions, the probabilities and the distribution with the associated parameters, have all been adjusted during the algorithm evolution.

As will become clear later from simulation results, with both optimized crossover and mutation functions, the updated GA shows significant performance gain over the basic GA.

\subsection{AutoPort: Direct Search of A Heuristic Algorithm}
 Different from the above strategy that  only modifies part of an algorithm, here we introduce the second strategy called EoH that utilizes LLM to
 design a complete heuristic algorithm without any prior restriction on the family of algorithms that can be used for port selection. To this end,
  we need to provide EoH more detailed instructions about the port selection task, which also involves solving the SINR balancing problem given a port selection and this function is given as sinr\_balancing().

 The programme template   is given in Listing \ref{lst:EoH:prompt} which provides the descriptions of the input and output parameters, and the specific task. It guides the LLM to search for an effective heuristic algorithm that solves the port selection problem for a block of $b$ channel realizations.

  \begin{lstlisting}[caption={The programme template for the AutoPort function.},captionpos=b, label={lst:EoH:prompt}]
  template_program = '''
def select_ports(K,n,N,Pt,B,H,noise_power):
    """
   Select n out of N ports to maximize the average objective value of B communications channels.

    Args:
        K: The number of users.
        n: The number of ports to be selected.
        N: Total number of ports available.
        Pt: Total transmit power available.
        b: Total number of channel realizations.
        H: Numpy array of shape (B,N,K). It denotes B channel realizations.
        noise_power: Noise power.
    Returns:
        port_sample: Numpy array of shape (B, n). For each row of it, all values should be integers from 0 to N-1 and cannot be repeated.

    For the n-th channel realization H(n), suppose port is a valid port selection solution, and then the effective channel becomes h_n = H[n,port,:]. The objective value will be calculated using the pre-defined function f_n=sinr_balancing(n, K, h_n, Pt, noise_power).
    """
'''
task_description = "Implement a function that selects a subset of ports for each channel realization to maximize the average SINR of B communications channels."
  \end{lstlisting}

 \begin{table}[h]
\centering
\caption{Typical Autoport algorithms attempted in the evolution.}
\label{table:autoport}
\begin{tabular}{ |l|p{6cm}| }  \hline
  No. & Description \\  \hline
  1 &  The new algorithm uses a greedy approach with local search, where ports are selected iteratively by adding the port that maximizes the immediate gain in SINR, followed by a local optimization step to swap ports for further improvement.  \\  \hline
  2 &  The new algorithm uses a simulated annealing approach with adaptive cooling schedule, where port selection is optimized through temperature-controlled probabilistic acceptance of new solutions, focusing on both exploration and exploitation. \\  \hline
  3 &  The new algorithm uses particle swarm optimization with adaptive velocity updates and neighborhood topologies to efficiently explore the port selection space while maintaining diversity and convergence. \\ \hline
  Chosen & The new algorithm uses a randomized beam search approach, where multiple candidate subsets are generated and refined in parallel, then the best subset is selected based on the objective value.  \\ \hline
\end{tabular}
\end{table}

Some typical algorithms in the evolution and the finally chosen algorithm are shown in Table \ref{table:autoport}. The code for the chosen AutoPort function in Python is depicted in Listing \ref{lst:EoH} of Appendix C. We can see that a number of heuristic methods including greedy algorithm, adaptive simulated annealing and particle swarm optimization have been attempted during the evolution.

The finally chosen AutoPort algorithm uses a greedy randomized adaptive search procedure (GRASP) \cite{GRASP}. The GRASP methodology was developed in the late 1980s
  for solving hard optimization problems.  GRASP is a multi-start metaheuristic algorithm consisting of two phases: a constructive randomized adaptive phase and a local search phase. In the first phase, the algorithm constructs multiple candidate solutions by randomly selecting ports. The sampling is not uniform but according to the probabilities weighted by channel gains of all ports.  Instead of choosing the best out of those greedy solutions, the GRASP algorithm improves each greedy solution in the second phase by applying a local search method to explore the local solution space. In the local search, the algorithm modifies one selected port of a greedy solution at a time, and evaluates the modified solution to check if it is better. The best of the local search solutions is returned as the final solution for the current starting solution. Applying local search starting from different greedy solutions may lead to a variety of different locally optimal solutions, and the best of which will  be  returned as the final output solution. We can see that the advantage of the GRASP algorithm is that it enables effective exploitation and exploration to  escape local optima via local search and multiple starts, and is thus well suited for solving the large-scale port selection problem.

\subsection{Complexity Analysis}




 The complexity of the overall design includes that for the evolutionary LLM-driven search and the complexity of the optimized algorithms.

 The time complexity for the evolutionary LLM-driven search is as follows.
\begin{itemize}
  \item Initialization (Step 1): The LLM generates $m$ initial heuristics. The complexity is $O(mT_{\text{LLM}})$ where $T_{\text{LLM}}$ is the complexity associated with access to the LLM.
  \item The Evolutionary Loop (Step 2): This step runs for $G$ generations. Within each generation, four sub-steps occur:
 \begin{itemize}
  \item Selection (Step 2.1): Sorting the population by fitness to assign ranks takes complexity of $O(m \log m)$. The complexity of the probabilistic selection for $m$ offspring is $O(m)$.
  \item Reproduction (Step 2.2): The LLM is queried to generate $m$ new offspring, so the complexity is $O(mT_{\text{LLM}})$.
  \item Evaluation (Step 2.3): This is typically the most computationally expensive part. Each of the $m$ new offspring must be evaluated, so the complexity is $O(m T_{\text{eval}})$, where $T_{\text{eval}}$ is the complexity when evaluating a batch of $b$ objective functions and will be analyzed later.
  \item Management (Step 2.4): Merging the old population and offspring ($2m$ total) and selecting the top $m$ requires another sorting, so the complexity is $O(m \log m)$.
\end{itemize}
\end{itemize}

 Considering all the above these steps, the total time complexity for LLM-driven algorithm search is $O(Gm(T_{\text{LLM}} + T_{\text{eval}} + \log m))$. Since $\log m$ is usually negligible compared to accessing the LLM inference and the batch evaluation of code, we can simplify this to   $O(Gm(T_{\text{LLM}} + T_{\text{eval}}))$.

The complexity of the optimized algorithms is dominated by the function evaluation, i.e., running the SINR-balancing algorithm [27] to find the optimal beamforming given a port selection. The algorithm used [27] includes the inversion of a matrix of $n\times n$ and the eigenvalue decomposition of a matrix of $K\times K$, so its complexity is $\mathcal{O}(n^3 + K^3)$ and $T_{\text{eval}} =\mathcal{O}(b(n^3 + K^3))$. The analysis of complexity of various optimized algorithms is as follows.
\begin{itemize}
  \item For the GA based algorithms, suppose the population size is $M$ and the number of iterations is $I$,  and then its computational complexity is $\mathcal{O}(MI(n^3 + K^3))$.
  \item For the GRASP algorithm obtained by AutoPort, in the worst case of each initial solution, $Nn$ evaluations are needed in the local search. Suppose there are $c$ initial solutions, then  its worst-case computational complexity is  $\mathcal{O}(Nnc(n^3 + K^3))$.
\end{itemize}

 \section{Simulation Results and Discussions}
 In this section, numerical experiments are conducted to evaluate and validate the proposed LLM-enabled optimization for port selection in a fluid antenna downlink MISO system.

 \subsection{Simulation Settings}
 The simulation setup follows the description in Section II.  Specifically, the system is operating with the carrier frequency $f_c = 2$ GHz,  with the BS transmit power of $20$ dBm,  the noise power spectral density of -174 dBm/Hz, and the bandwidth of 10 MHz. Unless otherwise specified, we consider the BS is equipped with a 2D FAS  with the port configuration $N_x=N_y=8$ and dimension $W_x = W_y=2$. The number of activated ports is the same as the number of users, i.e., $n=K$.

 We assume the distance between the BS and users is $d_k=200$ m, $\forall k$, and the large-scale path loss is given by
  \be
    \rho_k [dB] = 128.1+37.6 \log_{10}(d_k/1000).
 \ee
For each setting, we generate 1,000 channel data for LLM-enabled algorithm optimization and   100 data for the evaluation of all schemes.

 We use Deepseek-v3 as the default LLM. For the algorithm search, we employ $m=10$ initial heuristics, and $G=30$ generations. For the basic GA and optimized GA, $I=100$ and $50$ iterations are used, respectively. The population size is $M=20$ and $p=20\%$ of it are chosen as elite solutions. The maximum number of fitness evaluations is $F=300$. For the GRASP algorithm, $c=5$ initial solutions are considered.
We use a batch of $b=50$ channel realizations to optimize algorithms using EoH and during testing (inference), 100 channel realizations to evaluate the performance. Note that all LLM designed algorithms are optimized only once with the above  settings and $K=8$. The optimized algorithms are then generalized to other unseen settings during the evaluation.

\subsection{Evaluation and Comparison Schemes}
We evaluate   the performance of the following LLM-enabled schemes:
\begin{itemize}
  \item GA-C. This is the GA with EoH-optimized crossover operation.
  \item GA-CM. This is the GA with EoH-optimized crossover and mutation operations.
  \item AutoPort. This is the heuristic algorithm completely optimized by EoH.
\end{itemize}
We also compare them with the following benchmark schemes:
\begin{itemize}
  \item Exhaustive Search. It provides the optimal performance but is only possible when ${N \choose n}$ is not too large.
  \item Basic GA. This is the GA with basic crossover and mutation operations.
  \item  Radial Basis Function (RBF) method \cite{RBFOPT}. It is a representative traditional black-box optimization technique with open-source library and is highly competitive on  both continuous and mixed-integer nonlinear unconstrained problems.
  \item Transformer based solution \cite{transformer}\cite{transformer-Guo}: As a representative method of deep learning, we adopt a transformer with an eight-head attention module.
     It requires 30,000 training data.
  \item Random Selection. This scheme randomly chooses $n$ out of $N$ antenna ports.
\end{itemize}

\subsection{Performance Comparison}
\begin{figure}[]
	\centering
	\includegraphics[width=\linewidth]{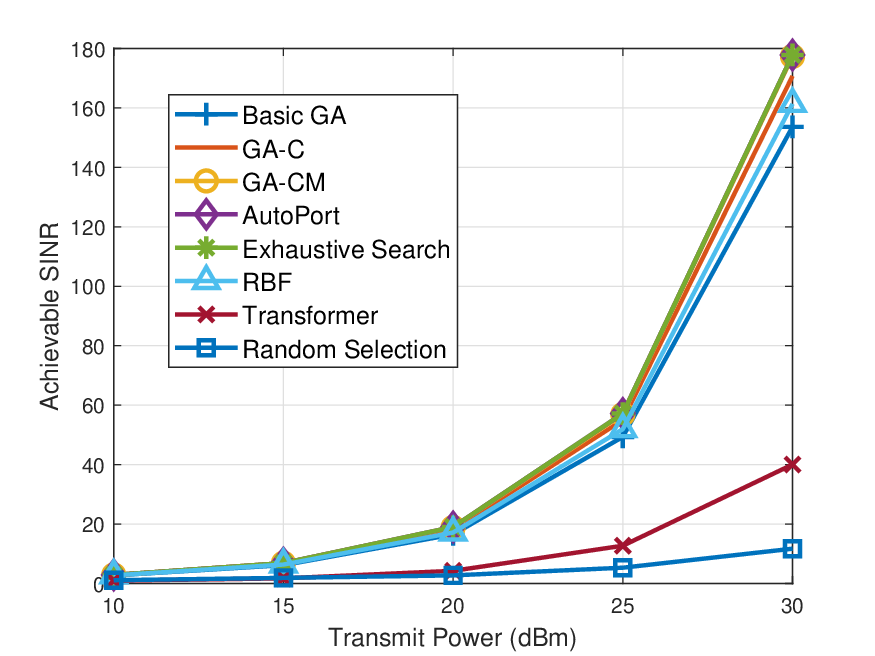}
	\vspace{-0.3cm}
	\caption{Achievable SINR vs the transmit power for $K=4$.}\label{fig:K4:power}
\end{figure}

We first consider a relatively small system which serves $K=4$ users. The achievable SINR is depicted in Fig. \ref{fig:K4:power} as the transmit power varies. We can observe that all three LLM-optimized algorithms achieve close to optimal performance.  The RBF method performs reasonably well and achieves near-optimal performance at low transmit power, but as the transmit power increases, its performance becomes worse.  The performance of the transformer and the random search is unsatisfactory.
The transformer also  exhibits limited performance that is even worse than the basic GA because it cannot effectively extract
features for the high-dimensional port selection problem. To better illustrate the performance difference, we also show the results in Table \ref{tab:K4:power} that is
normalized by the optimal solution using Exhaustive Search. We can clearly see that GA-C has significantly improved the performance of the basic GA by optimizing the crossover operation. By further optimizing the mutation operation, GA-CM achieves more than 99\% of the optimal performance. It is interesting to observe that AutoPort,   the scheme that is directly designed by  EoH, is even better than GA-CM, and can achieve the optimal performance.

 \begin{table}[]
\centering
\caption{Normalized SINR by   the exhaustive search vs  the transmit power, $K=4$.}
\label{tab:K4:power}
\begin{tabular}{ |l|l|l|l|l|l|  }
\hline
      $P$ (dBm)                    & 10    & 15     & 20      & 25 & 30   \\ \hline
  Basic GA        & 0.9458    & 0.8978   &  0.8678   &  0.8623    & 0.8640  \\ \hline
  GA-C     &  0.9769   & 0.9736   & 0.9512  &  0.9622  &  0.9602 \\ \hline
 GA-CM    &  0.9975 &   0.9940   & 0.9918  &  0.9941  &  0.9975    \\ \hline
 AutoPort  &  1.0000 &   1.0000   & 1.0000  &  1.0000  &  1.0000    \\ \hline
 RBF    &  0.9490  &  0.9276  &  0.9014 &   0.9080   & 0.9074    \\ \hline
 Transformer  & 0.3143   & 0.2387    &0.2080 &   0.2042  &  0.2062  \\ \hline
 Random    &  0.3843  &  0.2744  &  0.1433 &   0.0929  &  0.0659    \\ \hline
\end{tabular}
\end{table}

 \begin{table}[]
\centering
\caption{Normalized SINR by   the exhaustive search vs  the transmit power, $K=4$, Rician Channel.}
\label{tab:K4:power:Rician}
\begin{tabular}{ |l|l|l|l|l|l|  }
\hline
      $P$ (dBm)                    & 10    & 15     & 20      & 25 & 30   \\ \hline
  Basic GA        & 0.9794 &   0.9662   & 0.9490&    0.9426 &   0.9237 \\ \hline
  GA-C     & 0.9901  &  0.9825 &   0.9634  &  0.9654  &  0.9597\\ \hline
 GA-CM    & 0.9997  &  0.9990 &   0.9995  &  0.9975   & 0.9970   \\ \hline
 AutoPort  &  0.9999    &   1.0000   & 1.0000  &  1.0000  &  1.0000    \\ \hline
 RBF    &  0.9670   &  0.9471  &   0.9319 &    0.9173 &    0.8934   \\ \hline
  Transformer & 0.4776 &    0.3926 &    0.3251 &    0.3017  &   0.3034  \\ \hline
 Random    &  0.3248 &   0.2348   & 0.1362   & 0.0703  &  0.0316   \\ \hline
\end{tabular}
\end{table}

To further test the generalization ability of the proposed method, we extend the testing scenario where the small scale fading follows the Rician distribution with the Rician factor of 5 dB (note that
the default setting follows the Rayleigh distribution). The results are given in Table \ref{tab:K4:power:Rician}. We can see that similar to the results in Table \ref{tab:K4:power}, all proposed LLM-enabled approaches demonstrate excellent generalization performance compared to the RBF method, and AutoPort still achieves optimal or near-optimal performance. The performance of the transformer and random schemes is still unsatisfactory.
 \begin{table}[]
\centering
\caption{Running time of various algorithms (seconds), $K=4$.}
\label{tab:time}
\begin{tabular}{ |l|l|l|l|l|l|  }
\hline
      Basic GA     & GA-C      & GA-CM    & AutoPort & RBF  & Transformer\\ \hline
      1.29 &   1.28    & 1.35&   2.56 &   16.5 & 0.02  \\ \hline
 \end{tabular}
\end{table}
 Next we investigate the running time of various algorithms in Table \ref{tab:time}. All GA based algorithms require similar time, which shows the automated algorithm search does not increase the complexity of the original GA algorithm. {{AutoPort algorithm achieves superior performance, but requires about twice running time than GA based algorithms, so it is mainly suitable  for slow-varying fading channels}}. RBF has the highest complexity with medium performance. The transformer requires the least time but its performance is not satisfactory.

\begin{figure}[]
	\centering
	\includegraphics[width=\linewidth]{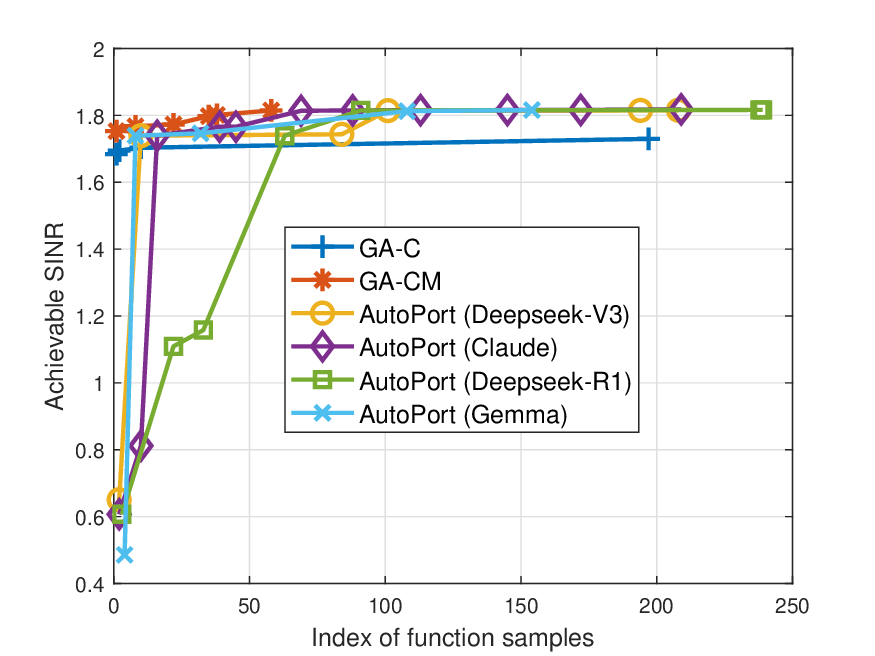}
	\vspace{-0.3cm}
	\caption{Convergence behaviour for $K=8$.}\label{fig:K8:convergence}
\end{figure}

In the rest of this section, we consider a system with $K=8$ users. Because of the excessively large number of parameters needed by the transformer, it is excluded for the comparison.
We first investigate the convergence performance of difference methods as shown in Fig. \ref{fig:K8:convergence}. For AutoPort, apart from Deepseek-V3, we have also tested other LLMs including Deepseek-R1,  a family of open reasoning models with performance approaching that of leading models, Claude 3.7 Sonnet,  the first hybrid reasoning model developed by Anthropic and Gemma3, one of the most capable models that runs on a single GPU. We can observe that all LLMs for AutoPort  converge fast (within 100 function samples) to similar SINRs, although Deepseek-R1 shows slower convergence. The GA-C and GA-CM methods converge even faster than AutoPort because they only optimize part of the GA.

\begin{figure}[]
	\centering
	\includegraphics[width=\linewidth]{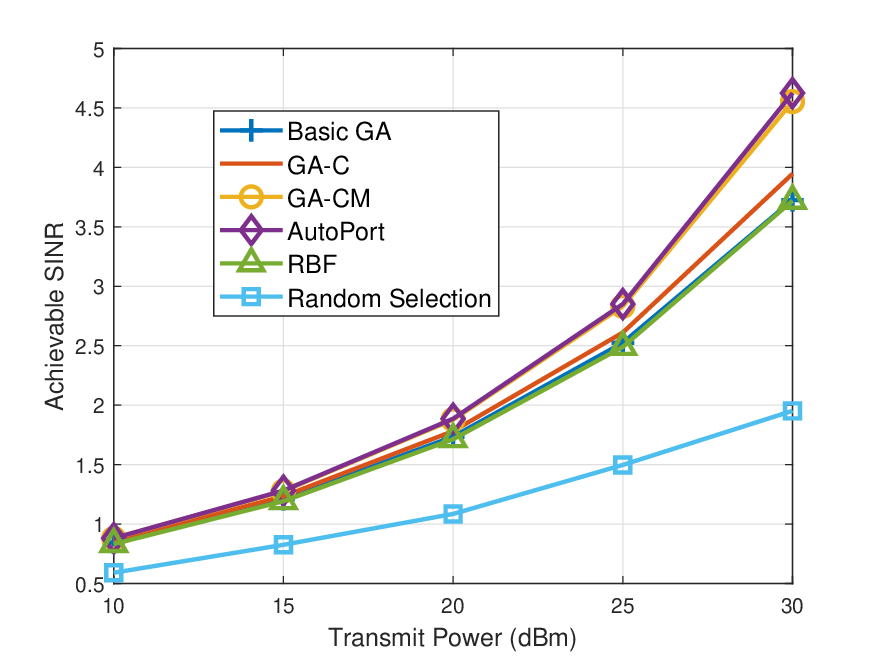}
	\vspace{-0.3cm}
	\caption{Achievable SINR vs the transmit power for $K=8$.}\label{fig:K8:power}
\end{figure}

Next we show the achievable SINR vs the transmit power in Fig. \ref{fig:K8:power}. It is observed that as the transmit power increases,
 LLM-optimized algorithms greatly outperforms the basic GA  and the random selection schemes. We normalize the achievable SINR of schemes by that of the basic GA and present the results in Table \ref{tab:K8:power}. A closer look at the results reveals that RBF achieves similar performance as the basic GA, while AutoPort achieves slightly better performance than GA-CM and nearly 1 dB gain over the basic GA scheme when the transmit power is 30 dBm.

 \begin{table}[h]
\centering
\caption{Normalized SINR by   the Basic GA   vs  the transmit power, $K=8$.}
\label{tab:K8:power}
\begin{tabular}{ |l|l|l|l|l|l|  }
\hline
      $P$ (dBm)                    & 10    & 15     & 20      & 25 & 30   \\ \hline
  GA-C     &  1.0157  &  1.0196 &   1.0269 &   1.0340  &  1.0594     \\ \hline
 GA-CM    &  1.0413   & 1.0555  &  1.0822  &  1.1216   & 1.2224   \\ \hline
 AutoPort  &  1.0425  &  1.0590 &   1.0870 &   1.1291  &  1.2416   \\ \hline
 RBF    &   0.9849 &    0.985&   0.9877  &   0.9863  &   0.9984 \\ \hline
 Random    &  0.6991  &  0.6823 &   0.6254 &   0.5931  &  0.5241    \\ \hline
\end{tabular}
\end{table}

\begin{figure}[]
	\centering
	\includegraphics[width=\linewidth]{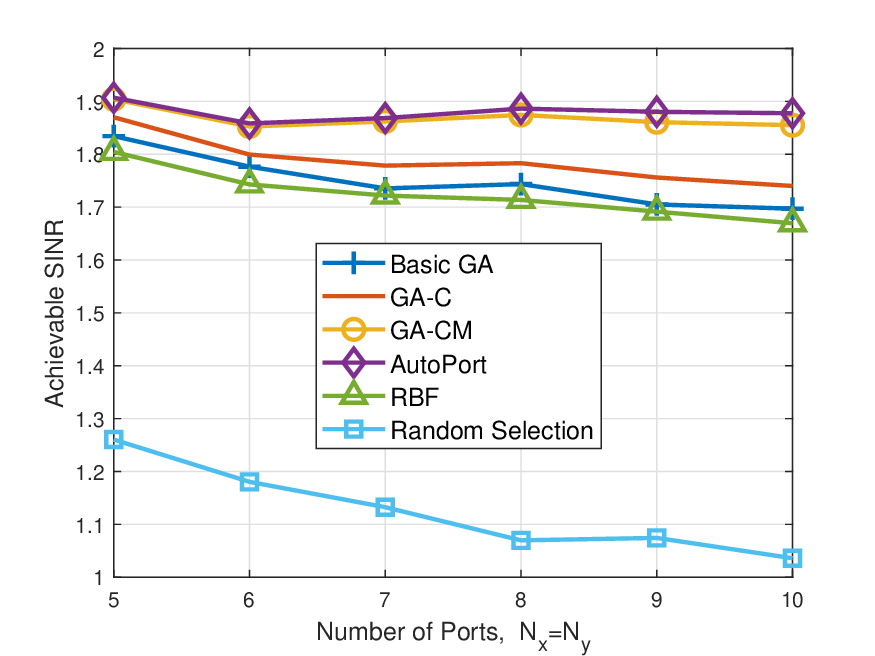}
	\vspace{-0.3cm}
	\caption{Achievable SINR vs the number of ports $N_x$ or $N_y$ for $K=8$.}\label{sinr_vs_port_K8}
\end{figure}

 We then investigate the impact of the number of ports on the achievable SINR with a fixed area of the 2D antenna panel,
 and results are shown in Fig. \ref{sinr_vs_port_K8}. As the number of ports increases, it is expected that the performance of all
 schemes will improve because of the increased spatial degrees of freedom. However, we can see that the achievable SINR of the random,
 basic GA, RBF and GA-C schemes all degrade with increasing number of ports. This can be explained by the fact  that  a higher number of ports, on the one hand
 increases the correlation between ports, and on the other hand increases the complexity of port selection, and therefore the resulting SINR of these three schemes becomes worse. By contrast, the performance of GA-CM and AutoPort improves as the number of ports increases and stabilizes when there are 64 or more ports,  which verifies their effectiveness.

\begin{figure}[]
	\centering
	\includegraphics[width=\linewidth]{ 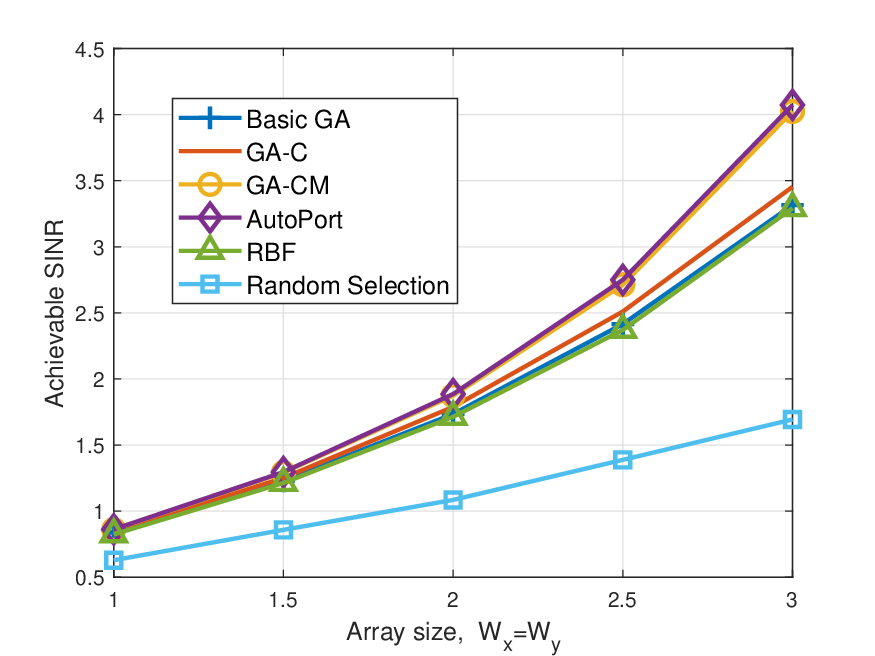}
	\vspace{-0.3cm}
	\caption{Achievable SINR vs the size of the 2D antenna array $W_x$ or $W_y$ for $K=8$.}\label{sinr_vs_length_K8}
\end{figure}

Finally, we evaluate the impact of size of the 2D antenna array $W_x$ or $W_y$ on the achievable SINR and results are given in Fig. \ref{sinr_vs_length_K8}.
It is seen that the achievable SINR improves as the array size increases. This is expected as large array size  reduces the correlation between ports and can better exploit the diversity gain. GA-C only achieves modest performance gain over the basic GA while with further optimized mutation, GA-CM outperforms the basic GA and RBF by more than 20\% when $W_x=3$ and approaches the performance of the AutoPort scheme. These results again shown the advantages of the GA-CM and the AutoPort schemes.

 \section{Conclusions}
 This paper presents an LLM-enabled automated algorithm design method for port selection in multiuser fluid-MIMO systems.
 Instead of directly optimizing the port selection,  this method designs   heuristic algorithms automatically to optimize the port selection using LLM.
 We have introduced two strategies of this method. One is to optimize the crossover and mutation operations in the GA while the other directly designs
 the complete heuristic algorithms. We have applied both strategies to the problem of maximizing the minimum SINR of all users. Through extensive simulations,
  the proposed method has shown superior performance over the basic GA for various system parameters and excellent generalization abilities,
   and therefore it provides a new direction for designing effective heuristic algorithms to solve difficult combinatorial problems in wireless
   optimizations.

   This paper is our first attempt towards automated algorithm discovery for wireless communications. We outline a few directions for future work. First, in the process of the proposed algorithm design, we have only considered performance but not complexity. How to jointly consider performance and complexity to achieve the desired balance is worth further study.
   Second, the channel estimation methods inherently introduce estimation errors. An in-depth study on how the inevitable CSI errors impact the robustness and performance of LLM-generated algorithms would be necessary. Third, in this paper, we have assumed a narrowband channel model to illustrate the potential of FAS and our proposed method. For
   wideband channel typically in 5G and 6G, FAS needs to be combined with wideband techniques like orthogonal frequency-division multiplexing (OFDM) \cite{FAS-OFDM} to deal with the intersymbol interference.

   In addition, apart from FAS, there are also other types of reconfigurable antenna systems such as the movable antenna \cite{movable2} that can change the antenna positions, and the rotatable antenna \cite{rotatable} that can adjust antenna orientation/boresight rotation without altering physical antenna positions, and  the six-dimensional (6D)  movable antenna \cite{movable1} capable of independently optimizing both  three-dimensional (3D) positions as well as 3D rotations within a given space. Our proposed automated algorithm design method holds great potential and can extend to optimize these new adjustable antenna parameters jointly with beamforming.

 \section*{Appendix}
 \subsection{Optimized Crossover Function}
  \begin{lstlisting}[caption={Optimized crossover function.},captionpos=b,label={lst:crossover}]
def crossover(parents, M):
    n_parents, n = parents.shape
    offspring = np.empty((M, n), dtype=parents.dtype)

    gene_freq = np.zeros((n, np.max(parents) + 1), dtype=float)
    for i in range(n):
        unique, counts = np.unique(parents[:, i], return_counts=True)
        gene_freq[i, unique] = counts / n_parents

    for i in range(M):
        p1, p2 = parents[np.random.choice(n_parents, 2, replace=False)]
        for j in range(n):
            if gene_freq[j, p1[j]] > gene_freq[j, p2[j]]:
                offspring[i, j] = p1[j]
            elif gene_freq[j, p1[j]] < gene_freq[j, p2[j]]:
                offspring[i, j] = p2[j]
            else:
                offspring[i, j] = p1[j] if np.random.rand() < 0.5 else p2[j]
    return offspring
\end{lstlisting}

\subsection{Optimized Mutation Function}
 \begin{lstlisting}[caption={Optimized mutation function.},captionpos=b, label={lst:mutate}]
def mutation(population, N):
    mutated_population = population.copy()
    p, n = population.shape

    for i in range(p):
        if np.random.random() < 0.7 and selected_port > 1 and np.random.random() < 0.5:
            idx1, idx2 = np.random.choice(n, size=2, replace=False)
            mutated_population[i, idx1], mutated_population[i, idx2] = mutated_population[i, idx2], mutated_population[i, idx1]

        num_mutations = max(1, int(selected_port * 0.15))
        idxs = np.random.choice(n, size=num_mutations, replace=False)
        noise = np.random.uniform(-N/5, N/5, num_mutations)
        mutated_population[i, idxs] = np.clip(
            np.round(mutated_population[i, idxs] + noise),
            0, N - 1
        ).astype(int)
    return mutated_population
\end{lstlisting}

\subsection{Optimized AutoPort Function}
 \begin{lstlisting}[caption={The optimized AutoPort function.},captionpos=b, label={lst:EoH}]
def select_ports(K, n, N, Pt, b, H, noise):
    port_sample = np.zeros((b, n), dtype=int)
    num_candidates = 5  # Number of candidate solutions to generate

    for i in range(b):
        best_rate = -np.inf
        best_ports = None
        port_norms = np.linalg.norm(H[i], axis=1)

        for _ in range(num_candidates):
            # Construct candidate solution with weighted random selection
            weights = port_norms / np.sum(port_norms)
            candidate = np.random.choice(N, size=n, replace=False, p=weights)

            # Local search improvement
            improved = True
            while improved:
                improved = False
                current_rate = sinr_balancing(n, K, H[i, candidate], Pt, noise_power)

                for j in range(n):
                    for alt_port in range(N):
                        if alt_port not in candidate:
                            temp_ports = candidate.copy()
                            temp_ports[j] = alt_port
                            temp_rate = sinr_balancing(n, K, H[i, temp_ports], Pt, noise_power)

                            if temp_rate > current_rate:
                                candidate = temp_ports.copy()
                                current_rate = temp_rate
                                improved = True

            if current_rate > best_rate:
                best_rate = current_rate
                best_ports = candidate.copy()

        port_sample[i] = best_ports

    return port_sample

  \end{lstlisting}

\end{document}